%% file: T-SP-17452-2014.tex
\documentclass[10pt,twocolumn,twoside]{IEEEtran}

\usepackage{amsmath}
\usepackage{amsfonts}
\usepackage{graphicx}
\usepackage{amssymb}
\usepackage{algorithmic,algorithm}
\usepackage{color}

\graphicspath{{./fig/}}

\newcommand{\undersmile}{\mathrel{\lower6pt\hbox{$\smile$}}}

\newcommand{\qd}{{\bf d}}

\newcommand{\qh}{{\bf h}}

\newcommand{\qn}{{\bf n}}

\newcommand{\qr}{{\bf r}}

\newcommand{\qv}{{\bf v}}
\newcommand{\qw}{{\bf w}}
\newcommand{\qx}{{\bf x}}

\newcommand{\qz}{{\bf z}}

\newcommand{\qB}{{\bf B}}

\newcommand{\qD}{{\bf D}}

\newcommand{\qH}{{\bf H}}
\newcommand{\qI}{{\bf I}}

\newcommand{\qR}{{\bf R}}

\newcommand{\qW}{{\bf W}}
\newcommand{\qX}{{\bf X}}

\newcommand{\tX}{{\tt X}}
\newcommand{\tA}{{\tt A}}
\newcommand{\tB}{{\tt B}}
\newcommand{\tR}{{\tt R}}

\newcommand{\qzero}{{\bf 0}}

\newcommand{\tr}{\textsf{trace}}

\newcommand{\be}{\begin{equation}} \newcommand{\ee}{\end{equation}}
\newcommand{\bea}{\begin{eqnarray}} \newcommand{\eea}{\end{eqnarray}}

\def\bl#1{{#1}}
\def\bbl#1{{blue}{#1}}

\begin{document}

\newcounter{MYtempeqncnt}

\title{Joint Beamforming Optimization and Power Control for Full-Duplex  MIMO  Two-way Relay  Channel}

\author{Gan Zheng, {\it Senior Member, IEEE}
\thanks{Gan Zheng is with School of Computer Science and Electronic Engineering, University of Essex, UK, E-mail: {\sf ganzheng@essex.ac.uk.}
 He is also affiliated with Interdisciplinary Centre for Security, Reliability and Trust (SnT),
  University of Luxembourg, Luxembourg.} }

\maketitle

\begin{abstract}
In this paper we explore the use of full-duplex radio  to improve
the spectrum efficiency in a two-way relay channel where two sources
  exchange information through an multi-antenna relay, and all
nodes work in the full-duplex mode. The full-duplex operation can
reduce the overall communication to only one phase but suffers from
the self-interference. Instead of purely suppressing the
self-interference, we aim to maximize the end-to-end performance by
jointly optimizing the beamforming matrix at the relay which uses
the amplify-and-forward protocol as well as the power control at the
sources. To be specific, we propose iterative algorithms and 1-D
search to solve two problems: finding the achievable rate region and
maximizing the sum rate. At each iteration, either the analytical
solution or convex formulation is obtained. We compare the proposed
full-duplex two-way relaying with the conventional half-duplex
two-way relaying, a full-duplex one-way relaying and a performance
upper bound. Numerical  results show that the proposed full-duplex
scheme significantly improves the achievable data rates over the
conventional scheme.
\end{abstract}


\section{Introduction}
 Cooperative communications via relaying is an effective measure to
 ensure reliability, provide high throughput and extend network
 coverage. It has been intensively studied in   LTE-Advanced \cite{4G-relay} and will continue to play an important role in
 the future fifth generation  wireless networks. In the conventional
 two-hop one-way relay channel (OWRC), the communication takes two orthogonal  phases, and  suffers from the loss of  spectral efficiency because of the
 inherent half-duplex (HD) constraint at the relay. Two-way relaying using the principle of physical layer network coding is  proposed to
 recover the loss in  OWRC by  allowing the two sources to exchange
 information more efficiently \cite{KRI}\cite{Zhang-2Phase}\cite{Qiang-Li-2009}. In the first phase of the two-way relay channel (TWRC), both sources
 transmit signals to the relay. In the second phase, the relay does not separate signals
 but rather broadcasts the processed mixed signals to both sources.
 Each source can subtract its own message from the received signal
 then decodes the information from the other source. The benefit of the TWRC is that,
 using the same two communication phases as the  OWRC, bi-directional communication is achieved.

 However, note that  the relay in the TWRC still operates in the HD mode thus two communication phases
 are needed.  Motivated by this observation and thanks to the emerging  full-duplex (FD) techniques,
 we aim to study the potential of the  FD operation
 in the TWRC to enable simultaneous information
 exchange between the two sources, i.e., only one communication phase is required.
 In the proposed scheme, all nodes, including the two sources
and the relay, work in the FD mode so they can transmit and receive
signals simultaneously \cite{BLISS}\cite{BLISS1}. However, the major
challenge in the FD operation is that the power of the
self-interference (SI) from the relay output could be over 100 dB
higher than that of the signal received from distance sources and
well exceeds the dynamic range of the analog-to-digital converter
\cite{RII3,RII2,DAY}. {\bl  Therefore it is important that the SI is
sufficiently mitigated.  The SI    cancellation  can be broadly
categorized as passive cancellation and active cancellation
\cite{Sabharwal-inband}. Passive suppression is  to isolate the
transmit and receive antennas using techniques such as directional
antennas, absorptive shielding and cross-polarization
\cite{Sabharwal-passive}. Active suppression is  to exploit a node's
knowledge of its own transmit signal to cancel the SI, which
typically includes analog cancellation, digital cancellation and
spatial cancellation. Experimental results are reported in
\cite{Duarte-Exp} that the SI can be  cancelled to make FD wireless
communication feasible in many cases. } In the recent work
\cite{Stanford-FD} and \cite{Stanford-FD-MIMO}, the promising
results show  that the SI can be suppressed to the noise level in
both single-antenna and multi-antenna cases. Below we will provide a
review on the application of the FD relaying to both the OWRC and
TWRC.

\subsection{Literature on the FD relaying}

\subsubsection{OWRC}
The FD operation has attracted lots of {\bl research interest}  for
relay assisted cooperative communication. It is shown in \cite{RII3}
that the FD relaying is feasible even in the presence of the SI and
can offer higher capacity than the HD mode. Multiple-input
multiple-output (MIMO) techniques provide an effective means to
suppress the SI in the spatial domain
\cite{RII1}-\cite{MIMO-relay-HD}. The authors of \cite{RII1} analyze
a wide range of SI mitigation measures when the relay has multiple
antennas, including natural isolation, time-domain cancellation and
spatial domain suppression. {  The FD relay selection is studied in
\cite{KRIKIDIS} for the amplify-and-forward (AF) cooperative
networks.} With multiple transmit or receive antennas at the
full-duplex relay, precoding at the transmitter and decoding at the
receiver can be jointly optimized to mitigate the SI effects. The
joint precoding and decoding design for the FD relaying is studied
in  \cite{RII1}, where both the zero forcing (ZF) solutions and the
minimum mean square error (MMSE) solutions are discussed.   When
only the relay has multiple antennas, a joint design of ZF precoding
and decoding vectors is proposed in \cite{CHOI_EL} to  null out the
SI at the relay. However, this design does not take into account the
end-to-end (e2e) performance. A gradient projection method  is
proposed  in \cite{Park-null-12}, to optimize  the precoding and
decoding vectors considering the e2e performance. When all terminals
have multiple antennas, the e2e performance is optimized in
\cite{MIMO-relay-HD} where the closed-form solution for
precoding/decoding vectors design as well as diversity analysis are
provided.

\subsubsection{TWRC}

 In the early work of TWRC, the FD operation is often employed to
 investigate   the capacity region from the  information-theoretic viewpoint without considering the effects of the SI \cite{capacity-twrc}\cite{capacity-twrc-half-bit}. 
 Only recently, the SI has been taken into account in the FD TWRC.
 In \cite{FD-two-phase}, the FD operation is introduced to the relay but two one-way relaying phases are required to achieve the two-way communication to avoid
 interference.
 A physical layer network coding  FD TWRC is proposed in \cite{FD-practical}, where bit error
rate is derived.  The optimal relay selection scheme is proposed and
analyzed in \cite{Song-FD-TW-Selection} for the FD TWRC using the AF
protocol. Transmit power optimization among the source nodes and the
relay node is studied   in \cite{FD-TW-SISO}, again using the AF
protocol. However, all existing works are restricted to the single
transmit/receive antenna case at the relay thus the potential of
using multiple antennas to suppress the SI and improve the e2e
performance for the TWRC is not fully explored yet.

\subsection{Our work and contribution}
 In this work, we study the potential of the MIMO FD operation
 in the TWRC where the relay has multiple transmit/receive antennas
 and employs the AF protocol and the principle of  physical layer network coding. The two sources have single transmit/receive antenna.
 We jointly optimize the relay beamforming matrix and the power
 allocation at the sources to maximize the e2e performance.
 To be  specific, we study two problems: one is to find the achievable rate
 region and the other is to maximize the sum rate of the FD TWRC.
 Our contributions are summarized as follows:
 \begin{itemize}
    \item We derive the signal model for the MIMO FD TWRC and
    propose to use the ZF constraint at the relay to greatly simplify the model
    and the problem formulations.
    \item We find the rate region by maximizing one source's rate
    with the constraint on the other source' minimum rate. We propose an
    iterative algorithm together with 1-D search to find the local
    optimum solution. At each iteration, we give analytical solutions for the transmit beamforming vector and the power control.
    \item  We tackle   the sum rate maximization problem by employing a similar iterative algorithm.
    At each iteration, we propose to use the DC (difference of convex functions) approach to optimize the transmit beamforming vector and we solve the power allocation analytically.
    \item We conduct intensive simulations to compare the proposed FD
    scheme with three benchmark schemes and clearly show its
    advantages of enlarging the rate region and improving the sum rate.
 \end{itemize}

The rest of the paper is organized as follows. In Section II, we
present the   system model, the explicit signal model and problem
formulations. In Section III, we deal with the problem of finding
the achievable rate region; in Section IV, we address the problem of
maximizing the sum rate. Three benchmark schemes are introduced in
Section V.  Simulation results are presented in Section VI. Section
VII concludes this paper and gives future directions.

 \noindent \emph{Notation}: The lowercase and
uppercase boldface letters (e.g., $\qx$ and $\qX$) indicate column
vectors and matrices, respectively. $\qX\in \mathbb{C}^{M\times N}$
means a complex matrix $\qX$ with dimension of $M\times N$.
 $\qI$ is the identity matrix.
 We use $(\cdot)^\dagger$ to denote the conjugate
transpose, $\tr(\cdot)$ is the trace operation, and $\|\cdot\|$ is
the Frobenius norm. $|\cdot|$ represents the absolution value of a
scalar. $\qX\succeq \qzero$ denotes that the Hermitian matrix $\qX$
is positive semidefinite. The expectation operator is denoted by
$\mathcal{E}(\cdot)$. Define $\Pi_\qX =
\qX(\qX^\dag\qX)^{-1}\qX^\dag$ as the orthogonal projection onto the
column space of $\qX$; and $\Pi_\qX^\bot = \qI - \Pi_\qX$ as the
orthogonal projection onto the orthogonal complement of the column
space of $\qX$.

\section{System model, signal model and problems statement}
\subsection{System model}\label{GENERAL_system_model}
Consider a  three-node MIMO relay network consisting of two sources
$\tA$ and $\tB$ who want to exchange information with the aid of a
MIMO relay $\tR$, as depicted in Fig. 1. {\bl There is no direct
link between the two sources because of deep fading or heavy
shadowing, so their communication must rely on $\tR$.} All nodes
work in the FD mode. To enable the FD operation, each source is
equipped with two groups of RF chains and corresponding antennas,
i.e., one for transmitting and one for receiving signals\footnote{It
is also possible to realize the FD operation using a single antenna,
see \cite{Stanford-FD}.}. We assume that each source has one
transmit antenna and one receive antenna. We use $M_{T}$ and $M_{R}$
to denote the number of transmit and receive antennas at $\tR$,
respectively. We assume $M_T>1$ or $M_R>1$ to help suppress the
residual SI at $\tR$ in the spatial domain.
\begin{figure}[t]
\centering
\includegraphics[width=0.9\linewidth]{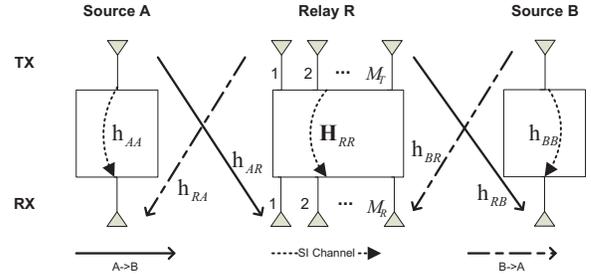}
  \caption{Full-duplex TWRC with two sources and a MIMO
relay. The dashed line denotes the  residual self-interference.}
\label{FIGG1}
\end{figure}
 We use $\qh_{XR}\in \mathbb{C}^{M_R\times 1}$ and
$\qh_{RX}\in\mathbb{C}^{M_T\times 1}$ to denote the directional
channel vectors between the source node $\tX$'s ($\tX\in
\{\tA,\tB\}$) transmit antenna to $\tR$'s receive antenna(s), and
between $\tR$'s transmit antenna(s) to $\tX$'s receive antenna,
respectively. In general, channel reciprocity does not hold, e.g.,
$\qh_{XR}\ne \qh_{RX}$, due to different transmit and receive
antennas used. In addition, $h_{AA}, h_{BB}$ and
$\qH_{RR}\in\mathbb{C}^{M_R\times M_T}$ denote the residual SI
channels   after the SI cancellation scheme is applied
 at the corresponding nodes \cite{KRIKIDIS,RII3}. {\bl The statistics of
the residual SI channel are not well understood yet \cite{SI-stat}.
The experimental-based study in \cite{Duarte-thesis} has
demonstrated that the amount of SI suppression achieved by a
combined analog/digital cancellation technique is a complicated
function of several system parameters. Therefore for simplicity, in
this paper, we model each element of the residual SI channel as a
Gaussian distributed random variable} with zero mean and variance
$\sigma^2_X, \tX\in \{\tA,\tB,\tR\}$. All channel links between two
nodes are subject to independent flat fading.  We assume that the
channel coefficients between different nodes remain constant within
a normalized time slot, but change independently from one slot to
another according to a complex Gaussian distribution with zero mean
and unit variance.  The global channel state information (CSI) is
available at the relay where the optimization will be performed. As
will be seen in Fig. \ref{fig:sumrate:csi}, this is critical for the
relay to adapt its beamforming matrix and for the two sources to
adjust their transmit power. The noise at the node $\tX$'s ($\tX\in
\{\tA,\tB, \tR\}$) receive antenna(s) is denoted as $n_X$ ($\qn_X$)
and modeled as complex additive white Gaussian noise with zero mean
and unit variance.

Since each source has only a single transmit and receive antenna, a
single data stream $S_A$ and $S_B$ are transmitted from $\tA$ and
$\tB$, with the average power $p_A$ and $p_B$, respectively. To keep
the complexity low, $\tR$ employs linear processing, i.e., the AF
protocol with an amplification matrix $\qW$,  to process the
received signal. The node $\tX$ has maximum transmit power
constraint $P_X$. We will jointly optimize the transmit power of the
two source nodes, $p_A$ and $p_B$ together with the amplification
matrix at the relay to maximize the e2e system performance.

 {\bbl The overhead due to the CSI acquisition at each node is analyzed as follows. The received CSI
$\qh_{AR}$ and $\qh_{BR}$ at  $\tR$ can be estimated  by $\tA$ and
$\tB$ each sending one pilot symbol separately. The SI channel
$\qh_{RR}$  can be estimated at $\tR$   by itself sending an
$M_T$-symbol pilot sequence. The SI channel $h_{AA}$ and $h_{BB}$
  are estimated similarly and then sent back to
$\tR$. Regarding the transmit CSI $\qh_{RA}$ and $\qh_{RB}$, $\tR$
first broadcasts an $M_T$-symbol  pilot sequence (this can  be used
for the estimation of $\qh_{RR}$ simultaneously), then $\tA$ and
$\tB$ feedback their   estimation to $\tR$. In addition, after $\tR$
performs the optimization and obtains $\qW$, $p_A$ and $p_B$, it
transmits $\qh_{RA}^\dag\qW \qh_{AR}$ and $\qh_{RB}^\dag\qW
\qh_{BR}$ to $\tA$ and $\tB$, respectively, such that they can
cancel their previously sent symbols. Finally, $\tR$ informs $\tA$
and $\tB$ their     transmit power   $p_A$ and $p_B$, respectively.}

\subsection{Signal model}
  We assume that the processing delay at $\tR$ is given by {\bbl a
 $\tau$-symbol duration}, which refers to the required
processing time in order to implement the FD operation \cite{RII1}.
$\tau$ typically takes integer values. The delay is short enough
compared to a time slot which has a large number of data symbols,
therefore its effect on the achievable rate is negligible.  At the
time ({\bbl symbol}) instance $n$, the received signal $\qr[n]$ and
the transmit signal $\qx_R[n]$ at $\tR$ can be written as
  \be\label{eqn:rn}
    \qr[n] = \qh_{AR}S_A[n] + \qh_{BR}S_B[n] + \qH_{RR} \qx_R[n] + \qn_R[n],
  \ee  \vspace{-3mm} and
  \be\label{eqn:xn}
    \qx_R[n] = \qW\qr[n-\tau],
  \ee
respectively. Using (\ref{eqn:rn}) and (\ref{eqn:xn}) recursively,
the overall relay output can be rewritten as \small
    \bea\label{eqn:xn2}
    \qx_R[n] &=& \qW \qh_{AR}S_A[n-\tau] + \qW\qh_{BR}S_B[n-\tau] \notag\\ &&+ \qW\qH_{RR} \qx_R[n-\tau] + \qW\qn_R[n-\tau]\notag\\
        &=&  \qW\sum_{j=0}^{\infty}\left(\qH_{RR}\qW\right)^j  \Big(\qh_{AR}S_A[n-j\tau-\tau] \notag\\ && + \qh_{BR}S_B[n-j\tau-\tau]+ \qn_R[n-j\tau-\tau] \Big),
  \eea  \normalsize
  {\bl where $j$ denotes the index of the delayed symbols}.

 Its covariance matrix is given by\small
  \bea
    &&{\mathcal{E}}[\qx_R\qx_R^\dag] = P_A \qW  \sum_{j=0}^{\infty}\left(\qH_{RR}\qW\right)^j \qh_{AR}\qh_{AR}^\dag
    \left( \left(\qH_{RR}\qW\right)^j\right)^\dag\qW^\dag \notag\\
    &&+P_B \qW  \sum_{j=0}^{\infty}\left(\qH_{RR}\qW\right)^j \qh_{BR}\qh_{BR}^\dag
    \left( \left(\qH_{RR}\qW\right)^j\right)^\dag\qW^\dag\notag\\
    && + \qW  \sum_{j=0}^{\infty} (\qH_{RR}\qW\qW^\dag \qH_{RR}^\dag )^j  \qW^\dag\notag\\
    &&= P_A \qW  \sum_{j=0}^{\infty}\left(\qH_{RR}\qW\right)^j \qh_{AR}\qh_{AR}^\dag
    \left( \left(\qH_{RR}\qW\right)^j\right)^\dag\qW^\dag \notag\\
    &&+ P_B \qW  \sum_{j=0}^{\infty}\left(\qH_{RR}\qW\right)^j \qh_{BR}\qh_{BR}^\dag
    \left( \left(\qH_{RR}\qW\right)^j\right)^\dag\qW^\dag\notag\\
    &&+\qW(\qI-\qH_{RR}\qW\qW^\dag\qH_{RR}^\dag)^{-1}\qW^\dag.
\eea\normalsize

 Note that $\tR$'s transmit signal covariance $\mathcal{E}[\qx_R\qx_R^\dag]$, and in turn the transmit power and the SI power, are  complicated functions of $\qW$,  which makes the optimization problems  difficult.
 To simplify the signal model and make the optimization problems more tractable, we add the   ZF constraint such that the optimization of $\qW$
 nulls out the residual SI  from the relay output to relay
input.  To realize this, it is easy to check from (\ref{eqn:xn2})
that the following condition is sufficient, \be
    \qW\qH_{RR}\qW=\qzero.
 \ee
 Consequently, (\ref{eqn:xn2}) becomes
    {\small\bea\label{eqn:xn3}
    \qx_R[n] =  \qW \left(\qh_{AR}S_A[n-\tau] + \qh_{BR}S_B[n-\tau] +
    \qn_R[n-\tau]\right),
  \eea}
  with the covariance matrix
  {\small
  \bea
    {\mathcal{E}}[\qx_R\qx_R^\dag] = p_A \qW\qh_{AR}\qh_{AR}^\dag\qW^\dag + p_B \qW\qh_{BR}\qh_{BR}^\dag\qW^\dag + \qW\qW^\dag.
\eea}
 The relay output power is
 \begin{align}
 p_R &=  \tr( {\mathcal{E}}[\qx_R\qx_R^\dag])\notag\\
 &= p_A\|\qW \qh_{AR}\|^2  + p_B\|\qW \qh_{BR}\|^2   +  \tr(\qW\qW^\dag).
 \end{align}

The received signal at the source  $\tA$ can be written as
  \bea
    r_A[n] &=& \qh_{RA}^\dag\qx_R[n] + h_{AA}S_A[n] + n_A[n]\notag\\
     &=& \qh_{RA}^\dag\qW \qh_{AR}S_A[n-\tau] +\qh_{RA}^\dag\qW  \qh_{BR}S_B[n-\tau] \notag\\&& + \qh_{RA}^\dag\qW\qn_R[n] + h_{AA}S_A[n] + n_A[n].
  \eea
After cancelling its own transmitted signal $S_A[n-\tau]$, it
becomes\footnote{Note that different from $S_A[n-\tau]$, $S_A[n]$
can not be completely cancelled due to the simultaneous
transmission, which is the main challenge of the FD radio.}
  \bea
    r_A[n] &=& \qh_{RA}^\dag\qW  \qh_{BR}S_B[n-\tau]  + \qh_{RA}^\dag\qW\qn_R[n]\notag\\
    && +  h_{AA}S_A[n] + n_A[n].
  \eea
 The received signal-to-interference-plus-noise ratio (SINR) at the source $\tA$, denoted as $\gamma_A$, is expressed as
  \bea\label{eqn:hi1}
    \gamma_A &=& \frac{p_B|\qh_{RA}^\dag\qW  \qh_{BR}|^2}{  \|\qh_{RA}^\dag\qW\|^2
    + p_A|h_{AA}|^2+     1}.
  \eea
  Similarly, the received SINR $\gamma_B$ at the source $\tB$ can be
  written as
  \bea\label{eqn:hi2}
    \gamma_B &=& \frac{p_A|\qh_{RB}^\dag\qW  \qh_{AR}|^2}{  \|\qh_{RB}^\dag\qW\|^2
    + p_B|h_{BB}|^2+ 1}.
  \eea
 The achievable rates are then given by $R_A = \log_2(1+\gamma_A)$ and $R_B =
 \log_2(1+\gamma_B)$, respectively.

\subsection{Problems Statement}
The conventional physical layer analog network coding scheme
requires two  phases for $\tA$ and $\tB$ to exchange information
\cite{Zhang-2Phase}.  Thanks to the FD operation, the proposed
scheme reduces the whole communication to only one phase thus
substantially increases the spectrum efficiency. However, the FD
operation  also brings the SI to each node so they may not always
use their maximum power because higher transmit power also increases
the level of the residual SI, therefore each node needs to carefully
choose its transmit power.

We are interested in two e2e objectives subject to each source's
power constraints by optimizing the relay beamforming and power
allocation at each source. The first one is to find the achievable
rate region $(R_A, R_B)$. This can be achieved by maximizing source
$\tA$'s rate while varying the constraint on the minimum rate of
source $\tB$'s rate (or vice versa), i.e.,
 solving the rate maximization problem $\mathbb{P}_1$ below:
 \bea\label{eqn:prob1}
    \mathbb{P}_1: \max_{\qW, p_A, p_B, p_R}{R_A} ~~ \mbox{s.t.}~~ R_B\ge r_B, p_X\le P_X,
    X\in\{A,B,R\},\notag
 \eea
 where $r_B$ is the constraint on source $\tB$'s rate. By enumerating
 $r_B$, we can find the boundary of the achievable rate region.

The second problem is to maximize the sum rate of the two sources.
Mathematically, this problem is formulated as $\mathbb{P}_2$ below:
 \bea\label{eqn:prob2}
    \mathbb{P}_2: \max_{\qW, p_A, p_B, p_R}{R_A+R_B} ~~ \mbox{s.t.}~~ p_X\le P_X,
    X\in\{A,B,R\}.\notag
 \eea
 The next two sections will be devoted to solving $\mathbb{P}_1$ and
 $\mathbb{P}_2$, respectively.

\section{Finding the achievable rate region}
In this section, we aim to optimize the relay beamforming matrix
$\qW$ and the sources' transmit power $(p_A, p_B)$  to find the
achievable rate region. This can be achieved by solving
$\mathbb{P}_1$. Using the monotonicity between the SINR and the
rate, it can be expanded as {\small
 \bea\label{eqn:fd2}
   &&   \max_{\qW, p_A, p_B} ~~ \frac{p_B|\qh_{RA}^\dag\qW  \qh_{BR}|^2}{  \|\qh_{RA}^\dag\qW\|^2
    + p_A|h_{AA}|^2+     1}\\
  &&   \mbox{s.t.} ~~  \frac{p_A|\qh_{RB}^\dag\qW  \qh_{AR}|^2}{  \|\qh_{RB}^\dag\qW\|^2
    + p_B|h_{BB}|^2+ 1} \ge \Gamma_B,\label{cons1}\\
    && p_A\|\qW \qh_{AR}\|^2  +  p_B\|\qW \qh_{BR}\|^2 +\tr(\qW\qW^\dag)  \le P_R,\label{cons2}\\
        && \qW\qH_{RR}\qW=\qzero,   \notag\\
    && 0\le p_A\le P_A, 0\le p_B\le P_B,\notag
 \eea}
 {\bl where $\Gamma_B\triangleq 2^{r_B}-1$ is the equivalent SINR constraint for the source $\tB$.}
 Observe that all terms in \eqref{eqn:fd2} are quadratic in $\qW$
 except the ZF constraint $\qW\qH_{RR}\qW=\qzero$, which is
 difficult to handle. Considering the fact that each source only transmits a single data
 stream and the network coding principle encourages mixing rather
 than separating the data streams from different sources,  we
 decompose $\qW$ as $\qW = \qw_t\qw_r^\dag$, where
$\qw_r$ is the receive beamforming vector and $\qw_t$ is the
transmit beamforming vector   at $\tR$. Then the ZF condition is
simplified to $(\qw_r^\dag\qH_{RR}\qw_t) \qW =\qzero$ or
equivalently $\qw_r^\dag\qH_{RR}\qw_t=0$ because in general $\qW\ne
\qzero$. Without loss of optimality, we further assume
$\|\qw_r\|=1$. As a result, the problem \eqref{eqn:fd2} can be
simplified to
   \bea\label{eqn:fd711}
    \max_{\qw_r, \qw_t, p_A, p_B} &&   \frac{p_B|\qh_{RA}^\dag \qw_t|^2|\qw_r^\dag  \qh_{BR}|^2}{  |\qh_{RA}^\dag \qw_t|^2 + p_A|h_{AA}|^2 +     1}  \\
    \mbox{s.t.} && \frac{p_A|\qh_{RB}^\dag \qw_t|^2 |\qw_r^\dag  \qh_{AR}|^2}{  |\qh_{RB}^\dag \qw_t|^2  + p_B|h_{BB}|^2+
     1} \ge \Gamma_B,\notag\\
     && p_A\|\qw_t\|^2 |\qw_r^\dag\qh_{AR}|^2  + p_B\|\qw_t\|^2 |\qw_r^\dag \qh_{BR}|^2  \notag\\ &&+ \|\qw_t\|^2 \le P_R,\notag\\
    && \qw_r^\dag\qH_{RR}\qw_t=0,   \notag\\
    && \|\qw_r\|=1,\notag\\
    && 0\le p_A\le P_A, 0\le p_B\le P_B.\notag
 \eea
 Note that in order to guarantee the feasibility of the ZF
 constraint $\qw_r^\dag\qH_{RR}\qw_t=0$, $\tR$ only needs to have
 two or more either transmit or receive antennas but not necessarily both.

 The problem \eqref{eqn:fd711} is still quite complicated as
 variables $\qw_t$, $\qw_r$ and $(p_A, p_B)$ are coupled. Our idea
 to tackle this difficulty is to use an alternating optimization
 approach, i.e., at each iteration, we optimize one variable while
 keeping the other fixed, together with 1-D search to find $\qw_r$. Details are given
 below.
\subsection{Parameterization of the receive beamforming vector $\qw_r$}
 Observe that $\qw_r$ is  mainly involved in $|\qw_r^\dag
\qh_{BR}|^2$ and
 $|\qw_r^\dag  \qh_{AR}|^2$, so it has to balance the signals received
 from the
 two sources. According to the result in \cite{Jorswieck2008f},
 $\qw_r$ can be parameterized by $0\le \alpha\le 1$ as below:
\be\label{eqn:wr} \qw_r =
\alpha\frac{\Pi_{\qh_{BR}}\qh_{AR}}{\|\Pi_{\qh_{BR}}\qh_{AR}\|} +
 \sqrt{1-\alpha}\frac{\Pi^\bot_{\qh_{BR}}\qh_{AR}}{\|\Pi^\bot_{\qh_{BR}}\qh_{AR}\|}.
 \ee
 {\bl We have to remark that \eqref{eqn:wr} is not the complete characterization of $\qw_r$ because it is also involved in the ZF constraint
 $\qw_r^\dag\qH_{RR}\qw_t=0$, but this parameterization makes the
 problem more tractable.}

 Given $\alpha$, we can  optimize $\qw_t$ and $(p_A,p_B)$ as will be introduced below. Then we perform 1-D search to find the
 optimal $\alpha^*$. We will focus on how to separately optimize
 $\qw_t$  and $(p_A, p_B)$ in the following
 two subsections. For the optimization of each variable, we will  derive analytical
 solutions without using iterative methods or numerical algorithms.

\subsection{Optimization of the transmit beamforming vector $\qw_t$}
 We first look into the  optimization of  $\qw_t$ assuming $\qw_r$ and $(p_A,
 p_B)$ are fixed. Based on the problem \eqref{eqn:fd711}, we get the
 following formulation:
   \bea\label{eqn:fd71123}
    \max_{\qw_t} &&   \frac{p_B|\qh_{RA}^\dag \qw_t|^2|\qw_r^\dag  \qh_{BR}|^2}{  |\qh_{RA}^\dag \qw_t|^2 + p_A|h_{AA}|^2 +     1}  \\
    \mbox{s.t.} && \frac{p_A|\qh_{RB}^\dag \qw_t|^2 |\qw_r^\dag  \qh_{AR}|^2}{  |\qh_{RB}^\dag \qw_t|^2  + p_B|h_{BB}|^2+
     1} \ge\Gamma_B,\notag\\
     && p_A\|\qw_t\|^2 |\qw_r^\dag\qh_{AR}|^2  + p_B\|\qw_t\|^2 |\qw_r^\dag \qh_{BR}|^2 \notag\\ &&
     + \|\qw_t\|^2  \le      P_R,\notag\\
    && \qw_r^\dag\qH_{RR}\qw_t=0.  \notag
 \eea
   By separating the variable $\qw_t$ and using monotonicity, \eqref{eqn:fd71123} is
 simplified to
   \bea\label{eqn:fd712}
    \max_{\qw_t} &&   |\qh_{RA}^\dag \qw_t|^2  \\
    \mbox{s.t.}
     &&  |\qh_{RB}^\dag \qw_t|^2  \ge\frac{\Gamma_B{p_B|h_{BB}|^2+1}}{p_A|\qw_r^\dag  \qh_{AR}|^2
     -\Gamma_B}\triangleq\bar\Gamma_B,
    \notag\\
     && \|\qw_t\|^2  \le   \frac{P_R}{(p_A|\qw_r^\dag\qh_{AR}|^2  + p_B  |\qw_r^\dag \qh_{BR}|^2  +1)}\triangleq\bar P,\notag\\
    && \qw_r^\dag\qH_{RR}\qw_t=0.  \notag
 \eea
  The problem \eqref{eqn:fd712} is not convex  but it is a
  quadratic problem in $\qw_t$.
  By defining $\qW_t = \qw_t
  \qw_t^\dag$ and using semidefinite programming relaxation,  \eqref{eqn:fd712} will become a convex problem in $\qW_t $, from which
 the optimal $\qw_t$ can be found from matrix decomposition. Interested readers are referred to \cite{SDPR} for details. However, the special structure of  the problem
 \eqref{eqn:fd712} allows us to derive the analytical solution in the following   steps.
\begin{enumerate}
    \item If $\bar\Gamma_B<0$ or $p_A|\qw_r^\dag  \qh_{AR}|^2 <\Gamma_B$, the
    problem is infeasible; otherwise continue.
    \item Define the null space of the vector $\qw_r^\dag\qH_{RR}$
    as $\mathbf{N}_t\in \mathbb{C}^{M_t \times (M_t-1)}$, i.e., $\qw_r^\dag\qH_{RR}\mathbf{N}_t=\qzero$. Introduce a new variable $\qv\in \mathbb{C}^{(M_t-1)\times 1}$ and express $\qw_t = \mathbf{N}_t\qv$, then
    we can remove the ZF constraint in \eqref{eqn:fd712}, and obtain
    the following equivalent problem:
   \bea\label{eqn:fd712e}
    \max_{\qv} &&   |\qh_{RA}^\dag \mathbf{N}_t\qv|^2  \\
    \mbox{s.t.}
     &&  |\qh_{RB}^\dag \mathbf{N}_t\qv|^2  \ge \bar\Gamma_B
    \label{eqn:constr1} \\
     && \|\qv\|^2  \le \bar P, \notag
 \eea
 where we have used the property that $\mathbf{N}_t^\dag
 \mathbf{N}_t=\qI$. If $\bar P\|\qh_{RB}^\dag \mathbf{N}_t\|^2< \bar\Gamma_B$, the problem
 is infeasible; otherwise continue.

 \item In this step we aim to  find the closed-form solution for \eqref{eqn:fd712e}. We first solve it without the constraint
 \eqref{eqn:constr1}.
 It can be seen that the last power constraint should always be
 satisfied with equality, and the optimal solution is given by
   $\qv^* = \sqrt{\bar P}\frac{\mathbf{N}_t^\dag \qh_{RA}}{\| \mathbf{N}_t^\dag
 \qh_{RA}\|}$. If it also satisfies the constraint \eqref{eqn:constr1},
 then it is the optimal solution; otherwise  continue.
 \item It this step, we know that both constraints in \eqref{eqn:fd712e} should be active, so we reach the
 problem below:
   \bea\label{eqn:fd713e}
    \max_{\qv} &&   |\qh_{RA}^\dag \mathbf{N}_t\qv|^2  \\
    \mbox{s.t.}
     &&  |\qh_{RB}^\dag \mathbf{N}_t\qv|^2  =\bar\Gamma_B
    \label{eqn:constr2}\notag\\
     && \|\qv\|^2  =\bar P. \notag
 \eea
\end{enumerate}
 If we define $\qd_2 \triangleq\frac{\mathbf{N}_t^\dag\qh_{RA}}{\mathbf{N}_t^\dag\qh_{RA}}, \qd_1\triangleq
 \frac{\mathbf{N}_t^\dag\qh_{RB}}{\mathbf{N}_t^\dag\qh_{RB}}$, $\phi\in (-\pi,\pi]$ be the argument of
 $\qd_2^\dag\qd_1$, $r\triangleq|\qd_2^\dag\qd_1|$, $q\triangleq \frac{\bar\Gamma_B}{\bar P\|\mathbf{N}_t^\dag\qh_{RB}\|^2}$ and   $\qz\triangleq\frac{\qv}{\|\qv\|}$, then we
 have the following formulation:
 \bea\label{eqn:lem5}
  \max_{\qz}&&  \qz^\dag \qd_2\qd_2^\dag\qz, \\
  \mathrm{s.t.}&& \qz^\dag \qd_1\qd_1^\dag\qz =   q, \quad \|\qz\|=1.\notag
 \eea

 The optimal solution $\qz^*$ follows from Lemma 2 in
\cite{Li-11} and is given below \bea\qz^*=
 \left( r\frac{1-q}{1-r^2}-\sqrt{q}\right) e^{j(\pi-\phi)}
 \qd_1+ \sqrt{\frac{1-q}{1-r^2}}\qd_2.
\eea
 Once we obtain $\qz^*$, the optimal transmit beamforming vector is
 given by
 \be \qw_t^* = \mathbf{N}_t \qv^* = \sqrt{\bar P}\mathbf{N}_t\qz^*. \ee

After obtaining the optimal $\qw_t^*$ using the above procedures, we
can move on to find the optimal power allocation at the   sources.

\subsection{Optimization of the source power $(p_A, p_B)$}
 Because of the FD operation at the sources and the fact that each source has a single transmit and receive antenna, they cannot suppress the residual SI in the spatial domain
 therefore cannot  always use the full power. In contrast, $\tR$ has at least two   transmit or receive antennas, so it can complete eliminate
 the SI and transmit using full power  $P_R$. Here we aim to  find the optimal power allocation $(p_A, p_B)$ at $\tA$ and $\tB$ assuming both $\qw_t$ and $\qw_r$ are
 fixed.

 For convenience, define $C_{At}\triangleq |\qh_{RA}^\dag \qw_t|^2,
C_{rB}\triangleq |\qw_r^\dag \qh_{BR}|^2, C_{Bt}\triangleq
|\qh_{RB}^\dag \qw_t|^2, C_{rA}\triangleq |\qw_r^\dag  \qh_{AR}|^2
$, then \eqref{eqn:fd711} becomes{\small
  \bea\label{eqn:fd77}
    \max_{p_A, p_B} &&   \frac{p_B C_{At}C_{rB}}{  C_{At}  + p_A|h_{AA}|^2+     1}  \\
    \mbox{s.t.} && \frac{p_A C_{Bt} C_{rA}}{  C_{Bt}   + p_B|h_{BB}|^2+
     1} \ge \Gamma_B\label{eqn:linear:SINR}\\
     && p_A\|\qw_t\|^2 C_{rA}  + p_B\|\qw_t\|^2 C_{rB}  + \|\qw_t\|^2\le P_R,\label{eqn:linear:power}\\
     && 0\le p_A\le P_A, 0\le p_B\le P_B.\notag
 \eea}
The problem \eqref{eqn:fd77} is a linear-fractional programming
problem, and can be converted to a linear programming problem
\cite[p. 151]{Boyd}. Again, thanks to its special structure, we can
derive its analytical solutions below step by step.
\begin{enumerate}
    \item First we check whether the constraint \eqref{eqn:linear:SINR} is feasible. If  $P_A C_{Bt} C_{rA} \le \Gamma_B$, then the problem is
    infeasible; otherwise, continue.

    \item Next we  solve \eqref{eqn:fd77} by  ignoring  the   constraint
    \eqref{eqn:linear:power}. It is easy to check that at the
    optimum, at least one source should achieve its  maximum power.
     The power allocation depends on two cases:
    \begin{enumerate}
    \item If $ P_A \ge  \frac{\Gamma_B(  C_{Bt} \|\qw_r\|^2 + P_B|h_{BB}|^2+
     1)}{C_{Bt} C_{rA} }  $, then $p_B=P_B$, $p_A = \frac{\Gamma_B(  C_{Bt} \|\qw_r\|^2 + P_B|h_{BB}|^2+
     1)}{C_{Bt} C_{rA} }$; otherwise,
    \item $p_A=P_A, p_B = \min\left(P_B, \frac{\frac{p_A C_{Bt} C_{rA}}{\Gamma_B} - 1 - C_{Bt}
    }{|h_{BB}|^2}\right)$.
    \end{enumerate}

    \item We check whether the above obtained solution satisfies the constraint \eqref{eqn:linear:SINR}. If it does, then it is the optimal solution. Otherwise
     the constraint  \eqref{eqn:linear:SINR} should be met with
     equality.

      \item The optimal power allocation is determined by the equation
      set below,
    \bea
        \left\{ \begin{array}{l}
              p_A C_{Bt} C_{rA} =\Gamma_B(  C_{Bt}   + p_B|h_{BB}|^2+
     1),  \\
             p_A\|\qw_t\|^2 C_{rA}  + p_B\|\qw_t\|^2 C_{rB}  + \|\qw_t\|^2
    =P_R
           \end{array}\right.
    \eea
    and the solution is given by
    \bea
        \left\{ \begin{array}{l}
                    p_A=     \frac{\Gamma_B|h_{BB}|^2}{C_{Bt}  + C_{rA}} p_B + \frac{\Gamma_B(C_{Bt}+1)}{C_{Bt}  +
                    C_{rA}},\\
                          p_B =  \frac{\frac{P_R}{\|\qw_t\|^2} -1 - \frac{\Gamma_B C_{rA}(C_{Bt}+1)
             }{C_{Bt}C_{rA}}}{\frac{\Gamma_B C_{rA}|h_{BB}|^2
             }{C_{Bt}C_{rA}} + C_{rB}  }.
           \end{array}\right.
    \eea
\end{enumerate}

\subsection{The overall algorithm}
 Given an $\alpha$ or $\qw_r$, we can iteratively optimize $\qw_t$  and  $(p_A,
 p_B)$ as above until convergence. The value of the objective function monotonically  increases as  the iteration goes thus converges to a local
 optimum. We can then conduct 1-D search over $0\le \alpha\le 1$ to find the
 optimal $\alpha^*$ or $\qw_r$. By enumerating source $\tB$'s requirement $r_B$, we can numerically
find the boundary of the achievable rate region.

 {\bl About the complexity, we remark that at each iteration, the solutions of $\qw_t$ and $(p_A,
 p_B)$ are given in simple closed-forms, so the associated complexity is low.}

\section{Maximizing the Sum Rate}
 In this section, we aim to maximize the sum rate of the proposed FD TWRC, i.e., to solve $\mathbb{P}_2$, which is rewritten
 below,
    \bea\label{prob:sum:rate:max}
    \max_{\qw_t, \qw_r p_A, p_B} &&   \log_2\left(1+ \frac{p_B C_{rB} |\qh_{RA}^\dag \qw_t|^2}{  |\qh_{RA}^\dag \qw_t|^2 + p_A|h_{AA}|^2 +
    1}\right)\\ &&
     +\log_2\left(1+  \frac{p_A C_{rA}|\qh_{RB}^\dag \qw_t|^2  }{  |\qh_{RB}^\dag \qw_t|^2  + p_B|h_{BB}|^2+
     1} \right)\notag \\
    \mbox{s.t.}    && \|\qw_t\|^2  \le \frac{P_R}{ p_A C_{rA} + p_BC_{rB}   + 1},\notag\\
    && \qw_r^\dag\qH_{RR}\qw_t=0.   \notag
 \eea
We will use the same characterization of \eqref{eqn:wr} to find the optimal $\qw_r$ via 1-D
 search.  We then    concentrate on alternatingly optimizing the transmit beamforming vector $\qw_t$  and the power
 allocation $(p_A, p_B)$.
 \subsection{Optimization of the transmit beamforming vector $\qw_t$}
  We first study how to optimize $\qw_t$ given $\qw_r$ and  $(p_A, p_B)$.
  For convenience, we define a semidefinite matrix  $\qW_t=\qw_t\qw_t^\dag$. Then the problem \eqref{prob:sum:rate:max} becomes
    \bea\label{prob:sum:rate:max2}
    \max_{ \qW_t\succeq \qzero } && F(\qW_t)    \\
    \mbox{s.t.}    &&\tr(\qW) \le \frac{P_R}{ p_A C_{rA} + p_BC_{rB}   + 1},\notag\\
    && \tr(\qW_t \qH_{RR}^\dag\qw_r\qw_r^\dag\qH_{RR})=0,
    \notag\\
    && \mbox{rank}(\qW_t)=1, \notag
 \eea
  where $F(\qW_t) \triangleq   \log_2\left(1+ \frac{p_B C_{rB} \tr(\qW_t\qh_{RA}\qh_{RA}^\dag)}{  \tr(\qW_t\qh_{RA}\qh_{RA}^\dag) + p_A|h_{AA}|^2 +     1}\right)
     +\log_2\left(1+  \frac{p_A C_{rA} \tr(\qW_t\qh_{RB}\qh_{RB}^\dag)  }{  \tr(\qW_t\qh_{RB}\qh_{RB}^\dag)  + p_B|h_{BB}|^2+
     1} \right).$ Clearly $F(\qW_t)$ is not a concave function thus (\ref{prob:sum:rate:max2}) is a cumbersome optimization problem.
 To tackle it, we propose to
use the DC   programming \cite{An} to find a local optimum point. To
this end,  we express   $F(\qW_t)$, as a difference of two concave
functions $f(\qW_t)$ and $g(\qW_t)$, i.e., {  \begin{align}
  & F(\qW_t)\notag\\
 &=  \log_2\left( (p_B C_{rB}+1) \tr(\qW_t\qh_{RA}\qh_{RA}^\dag)  +
p_A|h_{AA}|^2 + 1)
   \right) \notag\\
   & -  \log_2\left(\tr(\qW_t\qh_{RA}\qh_{RA}^\dag)  + p_A|h_{AA}|^2 + 1\right)\notag\\
   &  + \log_2\left( (p_A C_{rA}+1) \tr(\qW_t\qh_{RB}\qh_{RB}^\dag)  + p_B|h_{BB}|^2 + 1)
   \right)\notag\\
    & -  \log_2\left(\tr(\qW_t\qh_{RB}\qh_{RB}^\dag)  + p_B|h_{BB}|^2 +
    1\right)\notag\\
&\triangleq  f(\qW_t) - g(\qW_t) \end{align}}
 where  {\small \begin{align} f(\qW_t)
&\triangleq   \log_2\left( (p_B C_{rB}+1)
\tr(\qW_t\qh_{RA}\qh_{RA}^\dag)  + p_A|h_{AA}|^2 + 1)
   \right) \notag\\
   & +\log_2\left( (p_A C_{rA}+1) \tr(\qW_t\qh_{RB}\qh_{RB}^\dag)  + p_B|h_{BB}|^2 + 1)
   \right),\notag\\
g(\qW_t) &\triangleq   \log_2\left(\tr(\qW_t\qh_{RA}\qh_{RA}^\dag)  + p_A|h_{AA}|^2 + 1\right)\notag\\
&  +\log_2\left(\tr(\qW_t\qh_{RB}\qh_{RB}^\dag)  + p_B|h_{BB}|^2 +
    1\right).\notag \end{align}}
  $f(\qW_t)$ is a concave function while $g(\qW_t)$ is a convex
  function. The main idea is to approximate $g(\qW_t)$ by a linear
  function.  The linearization (first-order approximation) of $g(\qW_t) $ around the point $\qW_{t,k}$
 is given by{\small
  \bea  &&g_L(\qW_t;\qW_{t,k}) =
   \frac{1}{\ln(2)}\frac{\tr\left((\qW_{t}-\qW_{t,k})\qh_{RA}\qh_{RA}^\dag\right)}{\tr(\qW_{t,k}\qh_{RA}\qh_{RA}^\dag)  + p_A|h_{AA}|^2 +
   1} \notag\\
     &&+ \frac{1}{\ln(2)}\frac{\tr\left((\qW_{t}-\qW_{t,k})\qh_{RB}\qh_{RB}^\dag\right)}{\tr(\qW_{t,k}\qh_{RB}\qh_{RB}^\dag)  + p_B|h_{BB}|^2 +
   1} \notag\\
   &&+\log_2\left(\tr(\qW_{t,k}\qh_{RA}\qh_{RA}^\dag)  + p_A|h_{AA}|^2 +
   1\right)\notag\\
   &&+\log_2\left(\tr(\qW_{t,k}\qh_{RA}\qh_{RA}^\dag)  + p_A|h_{AA}|^2 +
   1\right).
\eea}  Then the DC programming is applied to sequentially solve the
following convex problem,  \bea
\qW_{t,k+1} &= &\arg \max_{\qW_t}\ f(\qW_t) - g_L(\qW_t; \qW_{t,k}) \label{DCP}\\
&&\mbox{s.t.}\quad  \tr(\qW_t) = \frac{P_R}{ p_A C_{rA} + p_BC_{rB}   + 1},\notag\\
    && \tr(\qW_t \qH_{RR}^\dag\qw_r\qw_r^\dag\qH_{RR})=0
    \notag.
\eea To summarize, the problem \eqref{prob:sum:rate:max2} can be
solved by i) choosing an initial point $\qW_t$; and ii) for $k=0, 1,
\cdots$, solving (\ref{DCP}) until the termination condition is met.
Notice that in \eqref{DCP} we have ignored the rank-1 constraint on
$\qW_t$. This constraint is guaranteed to be satisfied by the
results in \cite[Theorem 2]{Huang-11} when $M_t>2$, therefore the
decomposition of $\qW_t$ leads to the optimal solution $\qw_t^*$ for
\eqref{prob:sum:rate:max}. When $M_t=2$, the ZF constraint in the
problem \eqref{prob:sum:rate:max} can determine the direction of
$\qw_t$, i.e., $\qw_t=\sqrt{p_t}\mathbb{N}_t$ where $p_t$ is the
transmit power and $\mathbb{N}_t\in \mathbb{C}^{2\times 1}$
represents the null space of $\qw_r^\dag\qH_{RR}$. Therefore the
optimization of $\qw_t$ reduces to optimizing a scalar variable
$p_t$, which can be found by checking the stationary points of the
objective function in \eqref{prob:sum:rate:max} and the boundary
point without using the DC programming. The same applies to the
special case of $M_t=1$.

\subsection{Optimization of source power $(p_A, p_B)$}
With $\qw_t$ and $\qw_r$ fixed, the sum rate maximization problem
\eqref{prob:sum:rate:max} about power allocation can be written as
  \bea\label{eqn:fd777}
    \max_{p_A, p_B} &&   \log_2 \left(1+ \frac{p_B C_{At}C_{rB}}{  C_{At}  + p_A|h_{AA}|^2+     1}\right)\notag\\
    &&+\log_2 \left( 1+\frac{p_A C_{Bt} C_{rA}}{  C_{Bt}   + p_B|h_{BB}|^2+
     1}  \right)  \\
    \mbox{s.t.} && p_A  C_{rA}  + p_B  C_{rB}  + 1 \le \frac{P_R}{\|\qw_t\|^2},\label{eqn:constr3}\\
     && 0\le p_A\le P_A, 0\le p_B\le P_B\notag.
 \eea
 Note that when the first relay power constraint \eqref{eqn:constr3} is not tight, the problem is
 the same as the conventional power allocation among two interference links to maximize the sum
 rate, and the optimal power solution is known to be binary \cite{binary-power},
 i.e., the optimal power allocation $ (p_A^*, p_B^*)$ should satisfy
 \be
    (p_A^*, p_B^*)\in\{(0, P_B), (P_A,0), (P_A, P_B)\}.
 \ee

 Next we only focus on the case in which the constraint \eqref{eqn:constr3} is
 active, i.e., $p_A  C_{rA}  + p_B  C_{rB}  + 1
 =\frac{P_R}{\|\qw_t\|^2}$.
We then have
 \be
    p_A = \frac{\frac{P_R}{\|\qw_t\|^2-1}-p_B  C_{rB}}{C_{rA} }.
 \ee
 Because $0\le p_A\le P_A$, we can obtain the feasible range $[p_B^{\min}, p_B^{\max}]$ for $p_B$:
 \bea
     p_B^{\min} &=& \max\left(0, \frac{ {P_R}{\|\qw_t\|^2}-1 - C_{rA} P_A }{ C_{rB}}
    \right),\notag \\
     p_B^{\max} &=& \min \left(P_B, \frac{{P_R}{\|\qw_t\|^2}-1}{ C_{rB}}\right).
 \eea

 The objective function of \eqref{eqn:fd777} then becomes a function of $p_B$ only, i.e.,
 \bea
    y(p_B) &=& \log_2 (C_{At}  + p_A|h_{AA}|^2+     1+ p_B C_{At}C_{rB})\notag\\&& -\log_2(  C_{At}  + p_A|h_{AA}|^2+     1)\notag\\
    &&+\log_2 ( C_{Bt}   + p_B|h_{BB}|^2+   1+ p_A C_{Bt} C_{rA})\notag\\&& -\log_2(  C_{Bt}   + p_B|h_{BB}|^2+
    1),
 \eea
 and \eqref{eqn:fd777} reduces to a one-variable optimization, i.e.,
 \be
  \max_{p_B} ~~ y(p_B)~~~~ \mbox{s.t.}~~  p_B^{\min}\le p_B\le  p_B^{\max}.
 \ee

 Setting $\frac{\partial y(p_B)}{\partial p_B}=0$ leads to
 {\small \bea
&&  \frac{ C_{At}C_{rB} -\frac{  C_{rB}}{C_{rA} } |h_{AA}|^2 }{ C_{At}  + \frac{\frac{P_R}{\|\qw_t\|^2-1}  |h_{AA}|^2}{C_{rA}}+     1
+ p_B \left( C_{At}C_{rB} -\frac{  C_{rB}}{C_{rA} } |h_{AA}|^2\right) }  \\
&& + \frac{\frac{  C_{rB}}{C_{rA} } |h_{AA}|^2}{  C_{At}  +
\frac{\frac{P_R}{\|\qw_t\|^2-1}}{C_{rA} } |h_{AA}|^2 +1   -
p_B\frac{  C_{rB}}{C_{rA} } |h_{AA}|^2 }\notag\\
&& +\frac{  |h_{BB}|^2- \frac{ C_{rB}}{C_{rA} } C_{Bt}
    C_{rA} }{    C_{Bt}    + \frac{\frac{P_R}{\|\qw_t\|^2-1}}{C_{rA} } C_{Bt} C_{rA}+ 1+ p_B\left( |h_{BB}|^2- \frac{ C_{rB}}{C_{rA} } C_{Bt}
    C_{rA}\right)}\notag\\
&& - \frac{|h_{BB}|^2}{ C_{Bt}   + p_B|h_{BB}|^2+1}
  =0.\notag
 \eea}
This in turn  becomes a cubic (3-rd order) equation and all roots
can be found analytically. Suppose that the set of all positive root
within $(p_B^{\min}, p_B^{\max})$ is denoted as $\Psi$ which may
contain  0, 1 or 3 elements. In order to find the optimal $p_B^*$,
we need to compare the objective values  of all elements in the set
$\Psi\cup\{p_B^{\min}, p_B^{\max}\}$ and choose the one that results
in the maximum objective value.

{\bl We comment that the complexity of the overall algorithm to
maximize the sum rate is dominated by the optimization of $\qw_t$
and more specially, solving the problem \eqref{DCP}. Since
\eqref{DCP} is a semi-definite programming (SDP) problem with one
variable $\qW_t\in \mathbb{C}^{M_t\times M_t}$  and two constraints,
the worst-case complexity to solve it is
$\mathcal{O}(M_t^{4.5}\log(\frac{1}{\epsilon}))$ where $\epsilon$ is
the desired solution accuracy \cite{SDPR}.}

\section{Benchmark schemes}
In this section, we introduce three benchmark schemes that the
proposed FD network scheme can be compared with. The first one is
the conventional two-phase HD TWRC using the analog network coding,
which is known to outperform the three-phase and four-phase HD
schemes  to provide throughput gain\cite{practical_PLNC}; the second
one is a two-phase one-way FD scheme  and in each phase, the relay
works in the FD mode\cite{FD-two-phase}; the last one ignores the
residual SI channel at the relay thus provides a performance upper
bound that is useful to evaluate the proposed algorithms.

\subsection{Two-phase HD relaying using analog network coding}
 HD analog network coding is introduced in \cite{Zhang-2Phase} which takes
 two phases to complete information exchange between $\tA$ and
 $\tB$. In the first phase, both sources transmit to $\qR$ and in
 the second phase, the relay multiplies the received signal by a
 beamforming matrix $\qW$ then  broadcasts it to $\tA$ and
 $\tB$. Because there is no SI, every node can use its full power,
 so only $\qW$ needs to be optimized.  The achievable rate pair
 and the relay power consumption, are given by
 \bea
    R_A &=& \frac{1}{2}\log_2\left(1+ \frac{P_B|\qh_{RA}^\dag\qW  \qh_{BR}|^2}{  \|\qh_{RA}^\dag\qW\|^2
    +1}\right),\notag\\
    R_B &=& \frac{1}{2}\log_2\left(1+ \frac{P_A|\qh_{RB}^\dag\qW  \qh_{AR}|^2}{
\|\qh_{RB}^\dag\qW\|^2  +   1}\right),\notag\\
   p_R &=& \|\qW \qh_{AR}\|^2 P_A + \|\qW \qh_{BR}\|^2 P_B
   +\tr(\qW\qW^\dag).\notag
 \eea
 where the factor of $\frac{1}{2}$ is due to the two
 transmission phases used.
 Given the above rates and the power expression, problems
 $\mathbb{P}_1$ and $\mathbb{P}_2$ can be solved to find the
 achievable rate region and the maximum sum rate, respectively.
 Compared with this scheme, the proposed FD relaying can reduce the
 total communication phases to one, thus has the potential to
 improve the throughput.

\subsection{Two-phase one-way FD}
 Another scheme that we will compare with is the FD one-way relaying
 in which $\tR$ works in the FD mode while the two sources work in the
 HD mode. In this way, both sources can transmit with the maximum power. We use the same notation as the proposed scheme and the relay beamforming matrix is  $\qW=\qw_t\qw_r^\dag$.

 The achievable rate, relay power constraint, and zero residual SI constraint for source $\tA$ (direction: $\tB \rightarrow\tA$)  are, respectively,
 \bea
    &&R_A =\log_2\left(1+  \frac{P_B|\qh_{RA}^\dag \qw_t|^2|\qw_r^\dag  \qh_{BR}|^2}{  |\qh_{RA}^\dag \qw_t|^2   +     1} \right), \\
 &&p_B\|\qw_t\|^2 |\qw_r^\dag \qh_{BR}|^2  +
\|\qw_t\|^2\|\qw_r\|^2\le P_R, \\
 &&\qw_r^\dag\qH_{RR}\qw_t=0.\label{eqn:ZF}
 \eea

 In our previous work \cite{MIMO-relay-HD}, we have derived the closed-form expressions below for $R_A$
 depending on how the ZF constraint is   realized:
 \begin{enumerate}
\item
  Receive ZF. In this case, we assume $\qw_t=\qh_{RA}$ and choose
  $\qw_r$ to achieve \eqref{eqn:ZF}. We showed that the achievable
e2e received signal-to-noise ratio (SNR) can be expressed as
 \bea\label{eqn:SNR:R}
    \gamma_{RZF}
         =\frac{P_B \|\qD\qh_{BR}\|^2 P_R \|\qh_{RA}\|^2}
    {P_B \|\qD\qh_{BR}\|^2 + P_R\|\qh_{RA}\|^2 + 1},
  \eea
 where $\qD\triangleq \Pi^\bot_{\qH_{RR}\qh_{RA}}$. 

\item Transmit ZF. In this case, we assume $\qw_r=\qh_{BR}$ and choose
  $\qw_t$ to achieve \eqref{eqn:ZF}. We then reach the following
  achievable SNR:
\bea\label{eqn:SNR:t}
    \gamma_{TZF} = \frac{  P_B \|\qh_{BR}\|^2 P_R\|\qB\qh_{RA}\|^2}{
        {P_B\|\qh_{BR}\|^2  + {P_R}\|\qB\qh_{RS}\|^2+ 1 }},
\eea  where $\qB\triangleq \Pi_{\qH_{RR}^\dag\qh_{BR}}^\bot$.
\end{enumerate}
 $R_A$ is then determined by $R_A=\log_2(1+\max(\gamma_{RZF},
 \gamma_{TZF}))$. We can derive similar achieve rate $R_B$  for the source
 $\tB$.

 Note that $R_A$ and $R_B$  cannot be achieved simultaneously as it requires
 that each corresponding source occupies the whole transmission time.
 The boundary of the  rate region can be obtained by using time-sharing parameter $t\in[0,1]$, i.e., $(t R_A, (1-t)R_B)$.

\subsection{FD-Upper Bound}
 This scheme is the same as the proposed FD scheme except that we
 assume there is no SI at $\tR$, but we still consider the SI at the two sources, i.e., $\qH_{RR}=\qzero, |h_{AA}|>0, |h_{BB}|>0$.
 In this case the ZF constraint in \eqref{eqn:fd2} is not necessary.
 We remark that this scheme uses unrealistic assumption
 of $\qH_{RR}=\qzero$, so it is not a practical scheme but provides a
useful upper bound to evaluate the performance of the proposed
algorithms.

In the simulation results, we will label the above three benchmark
schemes as ``Two-phase HD'', ``Two-phase FD'' and ``Proposed
one-phase FD upper bound'', respectively.

\section{Numerical Results}\label{sec:simu}
In this section  we provide numerical results to  illustrate the
achievable rate region and the sum rate performance  of the proposed
FD two-way relaying scheme. We compare it with the above mentioned
three benchmark schemes. The simulation set-up follows the system
model in Section II. Unless otherwise specified, we assume that
there are $M_T=M_R=3$ antennas at $\tR$,  the average residual SI
channel gain is  $\sigma^2_A=\sigma^2_B=\sigma^2_R=-20$ dB, {\bl and
the per-node transmit SNR for both sources and the relay is
$P_A=P_B=P_R=10$ dB,  which are the power constraints in problem
formulations $\mathbb{P}_1$ and $\mathbb{P}_2$}. The results are
obtained using 100 independent channel realizations.

\input figures.tex

\subsection{Achievable rate region}
First we illustrate the achievable {\bl average} rate region for the
two sources $\tA$ and $\tB$ in Fig. \ref{fig:rate:region}. {\bl It
is seen that the two-phase FD scheme already greatly enlarges the
achievable data region of the conventional HD scheme. The proposed
one-phase FD scheme achieves significantly larger rate region  over
the two-phase FD and the conventional  HD schemes.} We observe that
there is still a noticeable gap between the proposed scheme and its
upper bound because the proposed  solution  is suboptimal.

\subsection{Sum rate performance}
We then investigate the effect  of the source transmit SNR $P_A$ and
$P_B$ ($P_A=P_B$) on the sum rate  shown in Fig.
\ref{fig:sumrate:R1020dB} when { $P_R=10$ dB (solid curves) and
$P_R=20$ dB (dashed curves), respectively}. {\bl We first consider
the case $P_R=10$ dB.}  As expected, the sum rate improves as the
source transmit SNR increase and the proposed one-phase FD schemes
clearly outperforms the   two benchmark schemes. When the source SNR
is above 15 dB, the sum rate of the proposed FD scheme saturates.
This is because the high transmit power results in high residual SI,
therefore increasing power budget does not necessarily improve the
performance. To illustrate the performance improvement in sum rate,
we show the sum rate gain  over the conventional two-phase HD scheme
in Fig. \ref{fig:sumrate:gain:R1020dB}. It is observed  that when
the source SNR is 10 dB, the proposed one-phase FD scheme and the
two-phase FD scheme can achieve the sum rate gain of  1.56 and 1.22,
respectively. Even the performance upper bound cannot achieve double
rate because of the residual SI at both sources. The rate gain in
general decreases as the source transmit SNR increases again because
of the residual SI therefore the sources need to carefully adjust
its transmit power.

The same trend is observed   when $P_R=20$ dB. In this case, a sum
rate gain of nearly 1.7 is recorded when the source transmit SNR is
10 dB. Another interesting observation is that   the performance of
the proposed scheme is very close to the upper bound. {  This is
because when the relay power is high, the e2e performance is limited
by the link from the source to the relay, rather than the relay to
the other source, therefore the residual SI at the relay has little
effect on the sum rate.}

The impact of the relay transmit SNR on the sum rate is shown in
Fig. \ref{fig:sumrate:RPower}. It is seen that when the relay SNR is
low, the proposed one-phase FD scheme achieves lower sum rate than
the two-phase FD scheme at low transmit SNRs then outperforms the
latter when the transmit SNR is above 5 dB. The performance gain is
remarkable when the  relay transmit SNR is high.  This is because
unlike the two sources,  the relay   can null out the residual SI
using multiple antennas, therefore it can always  use  the maximum
available power to improve the sum rate.

The effect of the   residual SI channel gain at the sources is
examined in Fig. \ref{fig:sumrate:SI}. Naturally, the sum rate
decreases as the residual SI channel becomes stronger or the SI is
not adequately suppressed. When the residual SI channel gain is
above -5 dB, the two-phase FD scheme outperforms the one-phase FD
scheme while both still achieve higher sum rate than the
conventional two-phase HD scheme  even when the SI channel gain is
as high as 5 dB.

 Next we show the sum rate and rate gain results when  the number of antennas ($M_R=M_T$) at the
relay varies from 2 to 6 in Fig. \ref{fig:sumrate:antennas} and
\ref{fig:sumrate:gain:antennas}, respectively. The sum rate steady
increases as more antennas are placed at the relay due to the array
gain. It is observed that the rate gain remains about 1.55 when the
number of antennas is greater than 2.

\subsection{Asymmetric channel gain}
 The above results are mainly for a symmetric case, i.e., both sources have
 similar power constraints and channel {\bl strengths}. Here we consider an
 asymmetric case where the average channel gain between $\tR$ and
 $\tB$ is -10 dB. We plot the rate region in Fig. \ref{fig:sumrate:region:asymmetric}
using the same system parameters as those in Fig. 2 except that the
gain of channel vectors $\qh_{BR}$ and
 $\qh_{RB}$ is 10 dB weaker. {\bl The results show that both sources'
 rates are reduced while the source $\tB$ suffers more rate loss.} This is because one
 source's channels to and from the relay will also affect the
 performance of the other source. The sum rate comparison is given
 in Fig. \ref{fig:sumrate:asymmetric} and it is observed the performance of the two-phase {\bl FD} scheme is very close to that of the proposed FD scheme at all SNR region. This can be explained
 by the fact that e2e performance is restricted by the channel quality between $\tR$
 and $\tB$, so the gain due to the simultaneous transmission of two sources  is limited.

{
\subsection{Impact of the local channel state information}
 Finally  we consider  the case that only the  receive CSI at each node is available but the transmit CSI is unknown. Because of  the lack of the transmit
    CSI,   the two sources use full power
 $P_A$ and $P_B$; $\qw_t$ at the relay is chosen arbitrarily to satisfy the ZF constraint and the relay power
 constraint. The sum rate performance is shown in Fig.
 \ref{fig:sumrate:csi}. It is seen that the proposed FD scheme still achieves significant performance gain over the HD relaying at low to medium transmit SNRs although all rates
 are much lower than the case with the global CSI in Fig. 3. Another
 notable difference is that at high transmit SNRs, the performance of the proposed FD scheme  degrades quickly.
 This is because the two sources need to adjust its transmit power
 rather than using full power. This highlights the importance of the
 global CSI for adapting  the transmit power and  adjusting the relay beamforming.
}

  \section{Conclusion  and Future work}
  We have investigated the application of the FD operation to MIMO TWRC, which requires only one phase for the two sources to exchange
  information. We studied two problems of finding the achievable
  rate region and maximizing the sum rate by optimizing the relay
  beamforming matrix and power allocation at the  sources.
  Iterative algorithms are proposed together with 1-D search to
  find the local optimum solutions. At each iteration, either
  analytical solution or convex formulation has been derived. We
  have conducted intensive simulations to illustrate the effects of
  different system parameters. The results show that the FD
  operation has great potential to achieve much higher data rates than the conventional HD TWRC.

{\bl Regarding the future directions, better suboptimal and the
optimal solutions are worth studying.} There are a couple of reasons
why the proposed algorithm is sub-optimal such as  the additional ZF
constraint, {\bl the incomplete characterization of the receive
beamforming vector}, and the alternating optimization algorithms.
Another direction is to study the use of multiple transmit/receive
antennas at the two sources. If a single data stream is transmitted,
the residual SI can be removed using the ZF criterion at the sources
as well. This actually simplifies the optimization as the two
sources can use the maximum power. However, multiple antennas can
support multiple and variable number of data streams, and when the
problem is coupled with the SI suppression, it will be much more
challenging. Thirdly, in this paper, we focus on the benefit of the
FD {\bl in terms of} spectrum efficiency. In \cite{practical_PLNC},
it is shown that for the HD case, three-phase transmission schemes
offers a better compromise between the sum rate and the bit error
rate than the two-phase scheme, especially in the asymmetric case.
It is worth investigating whether such an trade-off also exists for
the FD scenario.

\end{document}

%% file: figures.tex
\begin{figure}[!ht]
\centering
\includegraphics[width=0.8\linewidth]{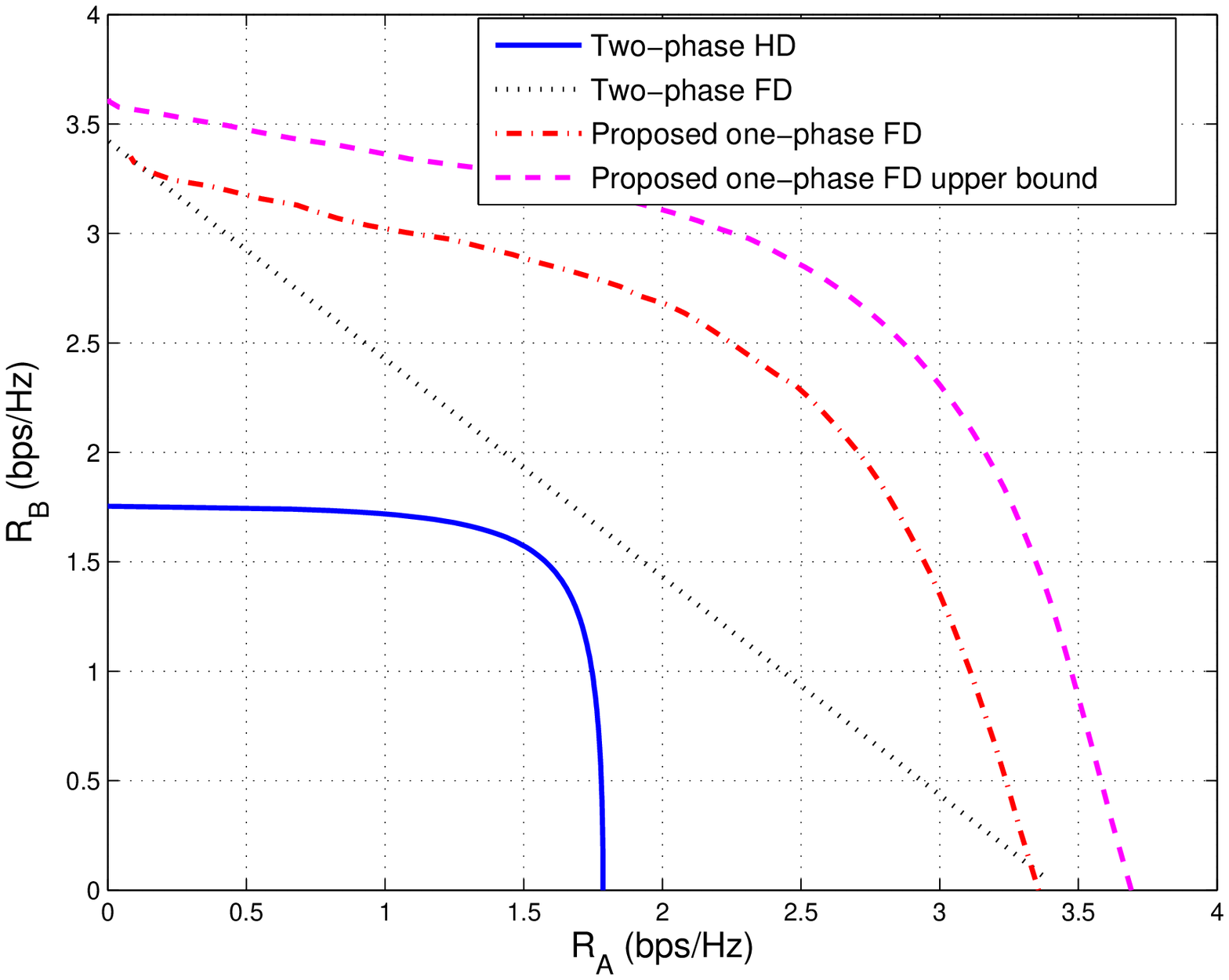}
\caption{Comparison of the achievable rate region. The residual SI
channel gain is -20 dB. Transmit SNR is 10 dB at all
nodes.}\label{fig:rate:region}\vspace{-5mm}
\end{figure}

\begin{figure}[!ht]
\centering
\includegraphics[width=0.8\linewidth]{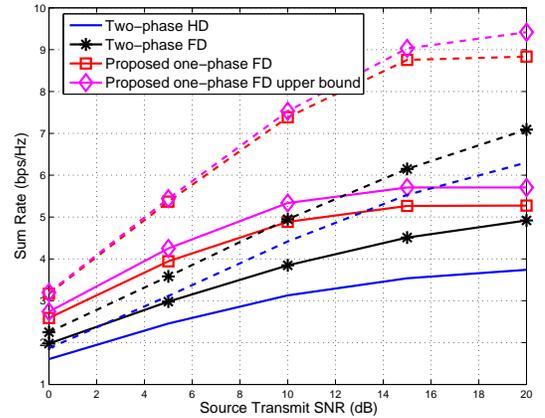}
\caption{Comparison of the sum rate. {Solid curves are for $P_R=10$
dB while dashed curves are for $P_R=20$ dB.}
}\label{fig:sumrate:R1020dB}
\end{figure}

\begin{figure}[!ht]
\centering
\includegraphics[width=0.8\linewidth]{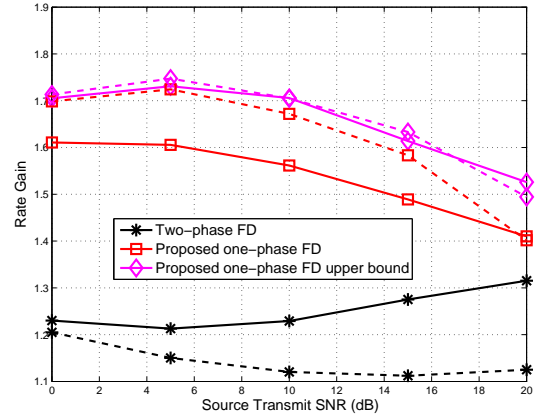}
\caption{Comparison of the sum rate gain over the conventional
two-phase HD scheme. {Solid curves are for $P_R=10$ dB while dashed
curves are for $P_R=20$ dB.} }\label{fig:sumrate:gain:R1020dB}
\end{figure}

\begin{figure}[!ht]
\centering
\includegraphics[width=0.8\linewidth]{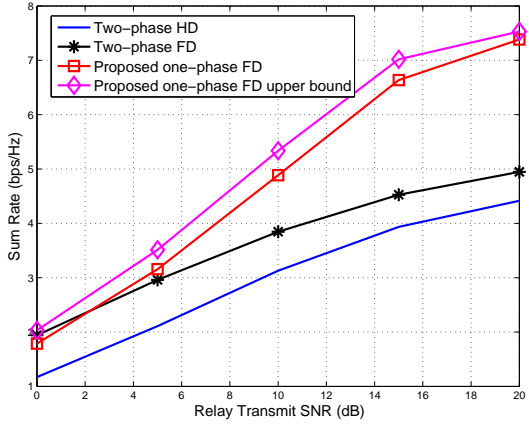}
\caption{The effect  of the relay transmit SNR on the sum rate,
$P_A=P_B=10$ dB.}\label{fig:sumrate:RPower}
\end{figure}

\begin{figure}[!ht]
\centering
\includegraphics[width=0.8\linewidth]{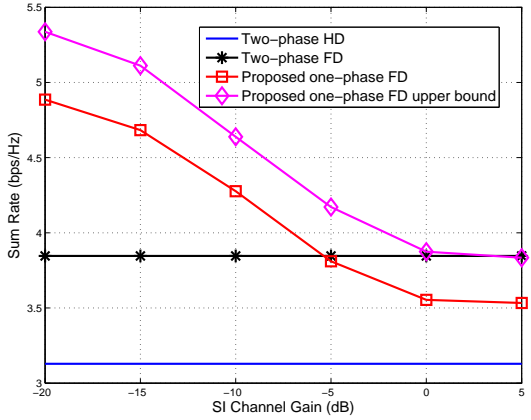}
\caption{The effect of the SI channel gain on the sum rate,
$P_A=P_B=10$ dB.}\label{fig:sumrate:SI}
\end{figure}

\begin{figure}[!ht]
\centering
\includegraphics[width=0.8\linewidth]{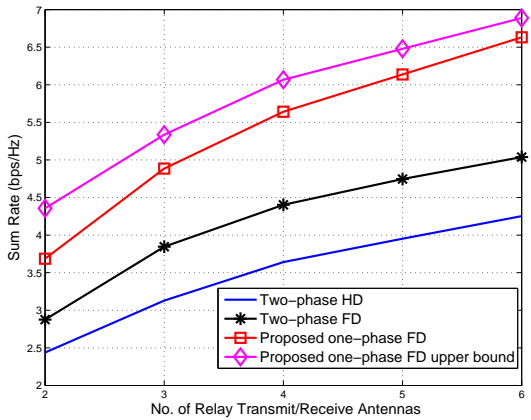}
\caption{The effect  of the number of the relay antennas on the sum
rate, $P_A=P_B=P_R=10$ dB.}\label{fig:sumrate:antennas}
\end{figure}

\begin{figure}[!ht]
\centering
\includegraphics[width=0.8\linewidth]{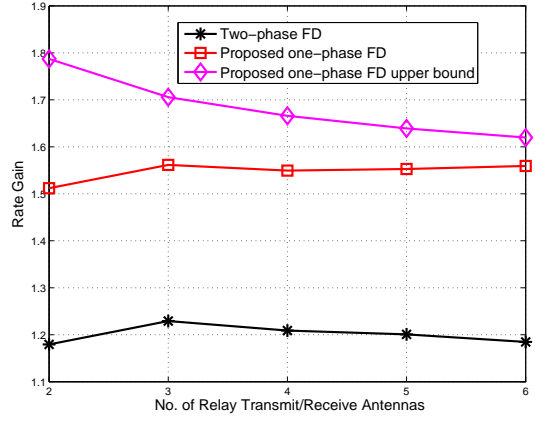}
\caption{Rate gain vs. no. of relay antennas, $P_A=P_B=P_R=10$
dB.}\label{fig:sumrate:gain:antennas}
\end{figure}

\begin{figure}[!ht]
\centering
\includegraphics[width=0.8\linewidth]{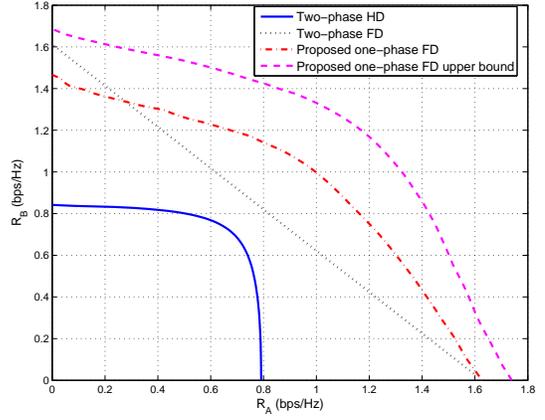}
\caption{Rate region for the asymmetric case, where the average
channel gain between $\tR$ and $\tB$ is -10
dB.}\label{fig:sumrate:region:asymmetric}
\end{figure}

\begin{figure}[!ht]
\centering
\includegraphics[width=0.82\linewidth]{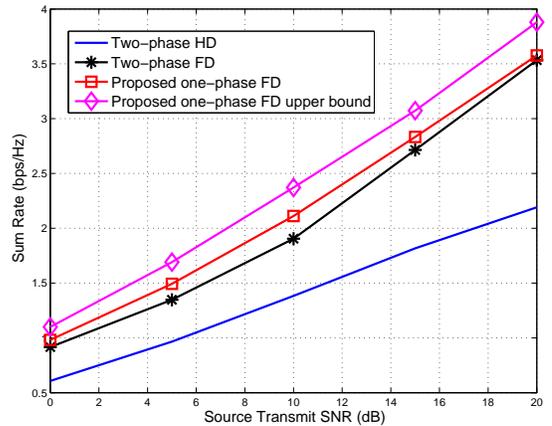}
\caption{Sum rate vs. source transmit SNR, $P_R=20$ dB for the
asymmetric case, where the average channel gain between $\tR$ and
$\tB$ is -10 dB.}\label{fig:sumrate:asymmetric}
\end{figure}

\begin{figure}[!ht]
\centering
\includegraphics[width=0.82\linewidth]{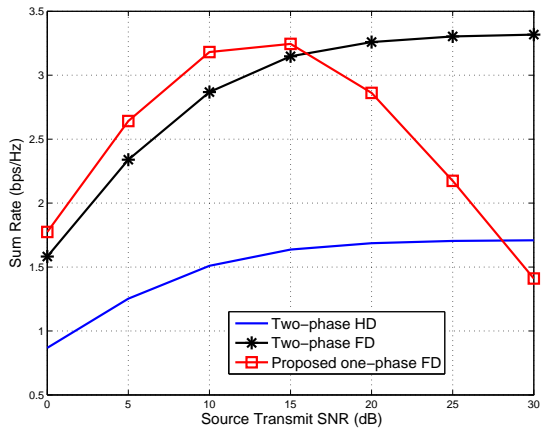}
\caption{Sum rate vs. source transmit SNR, $P_R=10$ dB with only the
receive CSI.}\label{fig:sumrate:csi}
\end{figure}